\documentclass[useAMS,usenatbib]{mn2e}
\usepackage{aas_macros}
\usepackage{graphicx}
\usepackage{mathptmx}
\usepackage{times}
\newcommand{\SNR}{\mbox{{G308.3--1.4}}}
\def\p0{\phantom{0}}

\begin{document}  
\title[Radio confirmation of Galactic supernova remnant \SNR]{Radio confirmation of Galactic supernova remnant \SNR}

\author[A. Y. De Horta et al. 2012]
{A.~Y.~De~Horta$^{1}$\thanks{E-mail: a.dehorta@uws.edu.au},
J.~D.~Collier$^{1}$,
M.~D.~Filipovi\'c$^{1}$,
E.~J.~Crawford$^{1}$, 
D.~Uro\v{s}evi\'c$^{2,3}$,
\newauthor F.~H.~Stootman$^{1}$
and N. F. H. Tothill$^{1}$
\\
$^{1}$University of Western Sydney, Locked Bag 1797, Penrith South, DC, NSW
1797, Australia\\
$^{2}$Department of Astronomy, Faculty of Mathematics, University of Belgrade, Studentski trg 16, 11000 Belgrade, Serbia\\
$^{3}$Isaac Newton Institute of Chile, Yugoslavia Branch \\
}

\date{Accepted 2012 October 6.  Received 2012 October 1; in original form 2011 December 14}

\pagerange{\pageref{firstpage}--\pageref{lastpage}} \pubyear{2012}

\maketitle

\label{firstpage}

%**********************************************************************************************************************************************************
\begin{abstract}
We present radio-continuum observations of the Galactic supernova remnant (SNR) candidate,  \SNR, made with the Australia Telescope Compact Array, Molonglo Observatory Synthesis Telescope and the Parkes radio-telescope. Our results combined with Chandra X-ray images confirm that \SNR\ is a {\em bona fide} SNR with a shell morphology. The SNR has average diameter of $D=34\pm19$~\mbox{pc}, radio spectral index of $\alpha = -0.68 \pm 0.16$ and linear polarisation of $10\pm1\%$; We estimate the SNR magnetic field $B\approx 29~\mu\mbox{G}$. Employing a $\Sigma-D$ relation, we estimate a distance to \SNR\ of $d=19\pm11~\mbox{kpc}$. The radio morphology, although complex, suggests a smaller size for the SNR than previously implied in an X-Ray study. These results imply that \SNR\ is a young to middle-aged SNR in the early adiabatic phase of evolution. 
\end{abstract}
%**********************************************************************************************************************************************************

%**********************************************************************************************************************************************************
%**********************************************************************************************************************************************************

%**********************************************************************************************************************************************************
\section{Introduction}
The chemical and physical state of the interstellar medium (ISM) is significantly influenced  by supernovae (SNe) and their resulting supernova remnants (SNRs). SNe and SNRs are important drivers of Galactic evolution \citep{1994ApJS...92..487T, 2006NuPhA.777..424N} as they can induce star formation by accelerating cosmic rays, heating and compressing nearby clouds and injecting  heavy elements into the ISM. 
According to the continuously updated, online catalogue created by \cite{2009BASI...37...45G}, some 274 SNRs are identified in our Galaxy. Understanding the SNR population and its effects on the structure of our Galaxy requires as complete a sample of confirmed SNRs as possible.

\SNR\ was initially classified as a SNR candidate amongst $\sim350$ others by \citet{Busser98}, when conducting a search for extended and unidentified X-ray sources in the \emph{ROSAT} All-Sky Survey (RASS) database. \citet{2002ASPC..271..391S} narrowed these to 14 likely SNR candidates of which \SNR\ was one. They noted that \SNR\ exhibited a complex radio morphology but that it coincided with the RASS Position Sensitive Proportional Counter hard X-ray data suggesting more strongly the SNR status of \SNR. 

\SNR\ appears in the source catalogue \citep[p.223]{1994ApJS...91..111W} of the PMN survey at RA(J2000)=$13^\rmn{h}41^\rmn{m}$03\fs7, Dec(J2000)=$-63\degr40\arcmin58\farcs0$ with a flux density of 89$\pm8$ mJy. It also appears\footnote{incorrectly listed as G308.4-1.4} in the MOST SNR catalogue (MSC) \citep{1996A&AS..118..329W} with a flux density of $S_\nu=800$ mJy, dimensions 14$\times$4 arcmins and surface brightness $1.3\times10^{-21}\mbox{ W m\(^{-2}\) Hz\(^{-1}\) sr\(^{-1}\)}$. The MSC comments on the possible SNR having several arcs, suggesting a double-ring morphology.

Here, we present a new analysis of unpublished archival radio-continuum observations of the SNR candidate \SNR\ made with the Parkes radio telescope, the Australia Telescope Compact Array (ATCA) and the Molonglo Observatory Synthesis Telescope (MOST) at wavelengths $\lambda=6, 13, 20 \mbox{ and } 36$~cm. Our work confirms the status of \SNR\ as a {\em bona fide} SNR and suggest a different interpretation of its morphology to that of \citet{2012ApJ...750....7H} that also identify \SNR\ as an SNR from an analysis of the \mbox{X-ray} emission. We make the first estimate of its distance, polarisation, measurement of its spectral index and estimate its integrated flux density. 
%**********************************************************************************************************************************************************

%**********************************************************************************************************************************************************
\section{Observations and Data reduction}
%**********************************************************************************************************************************************************
\subsection{Australia Telescope Compact Array}
From  the Australia Telescope Online Archive\footnote{http://atoa.atnf.csiro.au} (ATOA) we extracted ATCA observational data of \SNR\ (Project C992; PI W. Becker). These radio data are re-reduced and published here for the first time. \SNR\ was observed at 13~cm and 20~cm in 128~MHz continuum mode: 32 full-polarisation 4~MHz channels are spread over 128~MHz intervals centred on 2496~MHz and 1384~MHz respectively. The first observation took 4.26 hours on 1-2 September 2001 in the 6B configuration (baselines from 214 to 5969~m), but these data were adversely affected by storms and covered only a narrow range of parallactic angle.
The second observation took place on 11th January 2002. The source was observed for 9.18 hours over a total of 11.36 hours in the 750A configuration (baselines from 77 to 3750~m).
In both cases, PKS 1934-638 was observed for flux density and bandpass calibration and PKS 1329-665 for phase calibration. 

\begin{figure}
  \includegraphics[width=65mm,angle=-90]{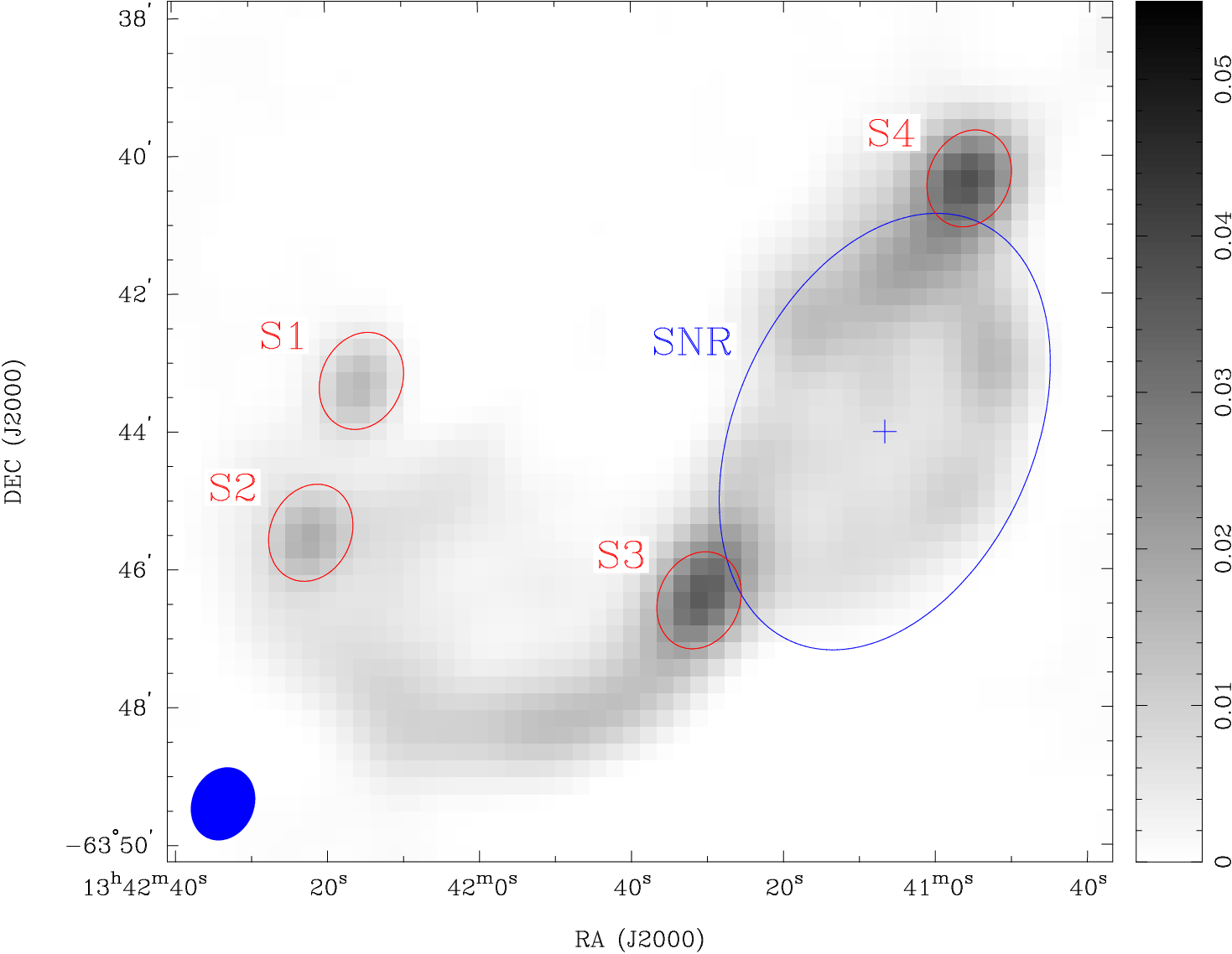}
  \caption{20~cm ATCA  image of Galactic SNR \SNR. 
The ellipse in the lower left corner represents the synthesised beam of \hbox{$65\farcs25\times53\farcs91$} at \hbox{P.A.$=-24\fdg1$}. Greyscale units are Jy/Beam.}
   \label{fig1}
\end{figure}

\begin{figure}
  \includegraphics[width=65mm,angle=-90]{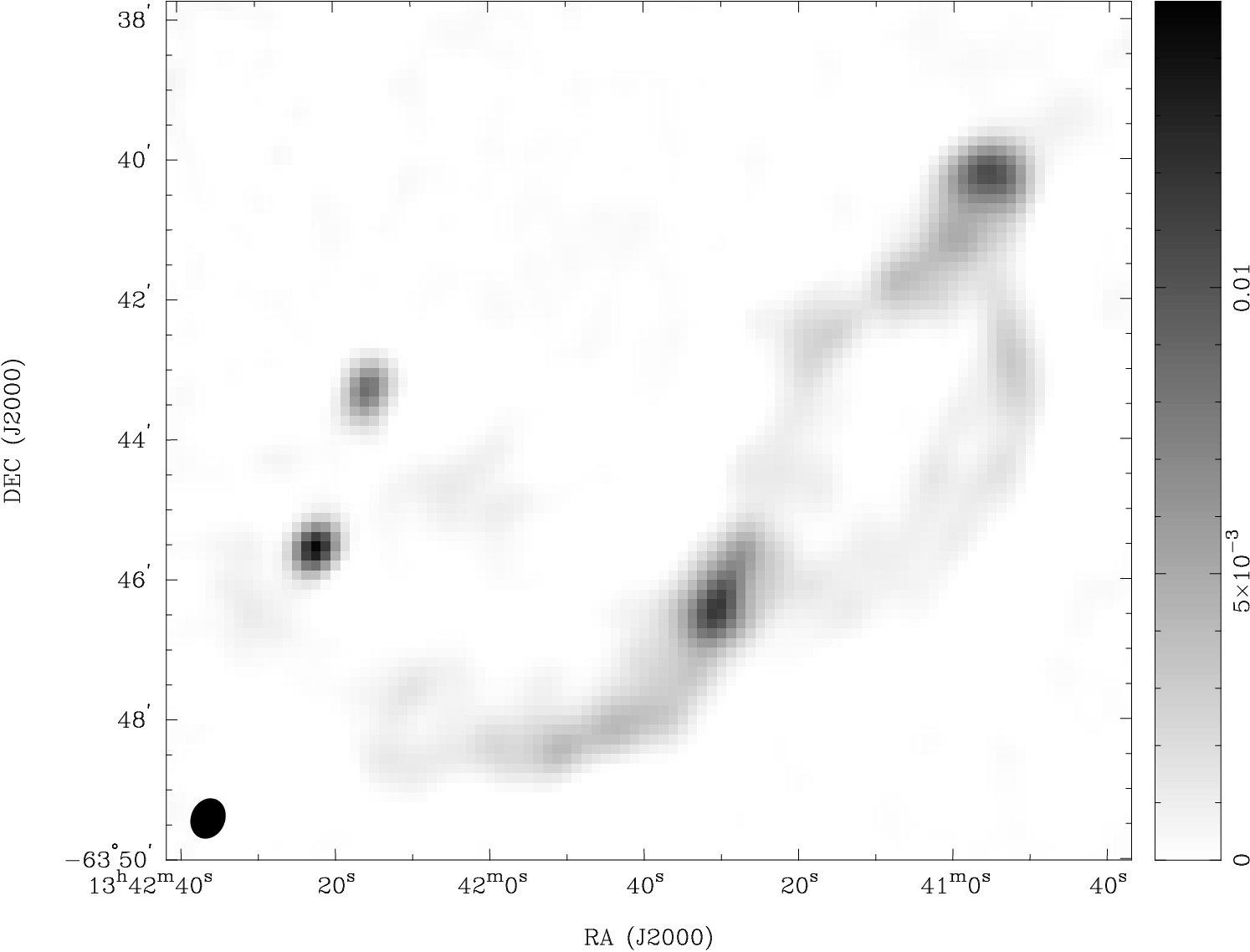}
  \caption{13~cm ATCA  image of Galactic SNR \SNR. 
The ellipse in the lower left corner represents the synthesised beam of \hbox{$35\farcs32\times28\farcs98$} at \hbox{P.A.$=-23\fdg5$}. Greyscale units are Jy/Beam.}
   \label{fig2}
\end{figure}

The \textsc{miriad} \citep{1995ASPC...77..433S} and \textsc{karma} \citep{2006karma} software packages were used for data reduction and analysis. The 6B data were not used: Combining them with the 750A data yielded no improvement of the final image, due to their poor parallactic angle coverage and contamination by severe interference.
The images were constructed using a natural weighting scheme, and the sixth antenna was removed to obtain the best signal to noise ratio, since the array was in a compact arrangement.
The final 20~cm image (Fig.~\ref{fig1}) has an estimated r.m.s noise level of 0.70~mJy~beam$^{-1}$, and
the final 13~cm image (Fig~\ref{fig2}) has 
an estimated r.m.s noise of 0.22~mJy~beam$^{-1}$.
These images suffer from missing short spacings and incomplete hour angle coverage, resulting in the highly negative regions seen enclosing the SNR emission. However, they are the highest-resolution images of this object currently available in the radio-continuum.

%FIGURE 1

%FIGURE 2
%**********************************************************************************************************************************************************

%**********************************************************************************************************************************************************
\subsection{Ancillary Data}
The MOST image we used was obtained from the Sydney University Molonglo Sky Survey (SUMSS) Postage Stamp Server (release 0.2)\footnote{http://www.astrop.physics.usyd.edu.au/cgi-bin/postage.pl}. SUMSS is a deep radio survey of the sky south of $\delta=-30^{\circ}$, covering $\sim3400$ square degrees. The catalogue \citep{2003MNRAS.342.1117M} contains $\sim100~000$ sources down to a flux density limit of 10 mJy~beam$^{-1}$ ($\sim$5$\sigma$). Each observation is a 12~hour synthesis, with a synthesised beamshape of $45\arcsec\times45\,\csc\delta\arcsec$ at position angle $0^\circ$.

The Parkes FITS image of \SNR\ used in this project was obtained from the Parkes-MIT-NRAO (PMN) 4850~MHz survey, downloaded from the ATNF database\footnote{http://www.parkes.atnf.csiro.au/observing/databases/pmn/pmn.html}. This  survey, which was conducted in June and November of 1990,  detected 36\,640 sources (see \cite{1993AJ....105.1666G} for the data reduction details). The final images have a 5\arcmin{} beam.

The {\em Chandra} satellite \citep{1996SPIE.2805....2W} observed \SNR\ on 26 June 2010 for 17\,ks (observation ID 11249), and we extracted these data products from the Chandra Data Archive (CDA)\footnote{http://cxc.harvard.edu/cda/}.

%**********************************************************************************************************************************************************

%**********************************************************************************************************************************************************
\section{Results}
%**********************************************************************************************************************************************************
\subsection{Morphology}

%FIGURE 3
\begin{figure}
  \hbox{\hspace*{-2mm}\includegraphics[width=65mm,angle=-90]{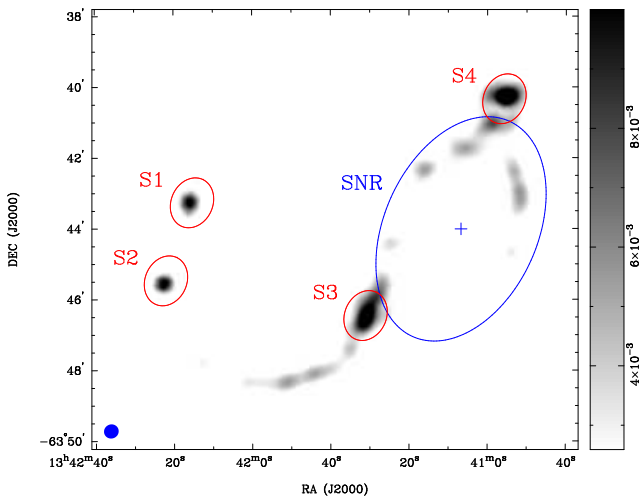}\hspace*{-100mm}}
  \caption{20~cm ATCA  image of Galactic SNR \SNR\ with sixth antenna included and clipped below \hbox{$10\sigma$}. The ellipse in the lower left corner represents the synthesised beam of \hbox{$25\farcs02\times23\farcs82$} at \hbox{P.A.$=-65\fdg08$}. \hbox{$\sigma$=25.9~mJy~beam$^{-1}$}. Greyscale units are Jy/Beam.}
   \label{fig3}
\end{figure}

%FIGURE 4
\begin{figure}
  \hbox{\hspace*{-2mm}\includegraphics[width=65mm,angle=-90]{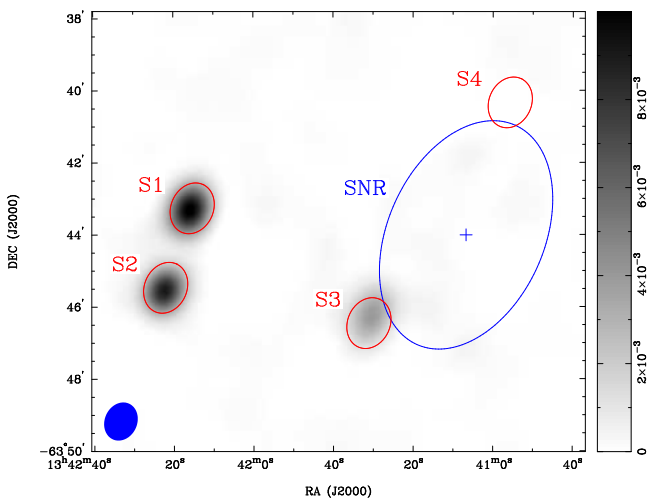}\hspace*{-100mm}}
  \caption{20~cm ATCA  image of Galactic SNR \SNR\ using only the sixth antenna, smoothed to the beam size and shape of Fig.~\ref{fig1}.  The ellipse in the lower left corner represents the synthesised beam of \hbox{$65\farcs25\times53\farcs91$} at \hbox{P.A.$=-24\fdg1$}. \hbox{$\sigma$=10.04~mJy~beam$^{-1}$}. Greyscale units are Jy/Beam.}
   \label{fig4}
\end{figure}

Figure~\ref{fig1} shows the elliptical region that we contend corresponds to SNR \SNR. We define  the SNR centre to be RA(J2000)=$13^\rmn{h}41^\rmn{m}$06\fs72, Dec(J2000)=$-63\degr44\arcmin01\farcs2$ (indicated by the cross in Fig.~\ref{fig1}), which corresponds to the mid point of this elliptical region. 

In considering the complex radio-continuum emission shown in Fig.~\ref{fig1} we need to determine how likely the point like emission is associated to the extended emission. The emission will be either intrinsic to \SNR\ or from unassociated foreground or background sources.

The bright regions on the north-west and south-east edges of this ellipse (S4 and S3 in Figs.~\ref{fig1} \& \ref{fig3}), in addition to the two sources seen east of the remnant (S1 and S2 in Figs.~\ref{fig1}, \ref{fig3} \& \ref{fig4}) do not have X-ray counterparts. We argue that these sources are not likely to be associated with the SNR in any way, and are most likely to be background sources (radio galaxy or AGN), or unresolved foreground sources. To confirm this we created a 20\,cm image (Fig~\ref{fig3}) including all baselines, ie. with the sixth antenna included, and with the Brigg's visibility weighting robustness parameter set to 0.5 in order to achieve the best balance between sidelobe suppression and signal to noise ratio. In this image S1 and S2 appear clearly as point sources, elongated at the position angle of the beam and with similar dimensions. Furthermore, S2 has an Infrared counterpart given in the InfraRed Astronomy Satellite (IRAS) Point Source Catalog  as IRAS 13388-6330. The source labelled S3 is a pointlike source with what appear to be radio jets coming out of either side of the centre to the north-west and south-east. To further test the pointlike properties of these sources another 20\,cm image (Fig~\ref{fig4}) was created using only the longest baselines, i.e.~those from antenna 6 to the other elements. S1, S2 and S3 retain their pointlike appearance, which further strengthens the case for them being background sources and not part of the SNR. We calculated 3-point spectral indices, $\alpha$\footnote{Spectral index defined as $S\propto\nu^\alpha$}, at wavelengths $\lambda$ = 36, 20 and 13~cm for sources S1 to S4 to be respectively: \mbox{$\alpha=-0.85\pm0.50$}; \mbox{$\alpha=-0.06\pm0.36$}, \mbox{$\alpha=-1.47\pm0.02$} and \mbox{$\alpha=-0.99\pm0.17$}.
The flat spectral index of S2 might indicate that it is a Galactic or foreground source, such as a compact HII region. However, we do not believe it to be associated with \SNR, as it is well outside the shock fronts.
Although source S4 and the SNR are resolved out in Fig.~\ref{fig4} (using the 5969 m ATCA baseline), S4 is still likely to be a radio galaxy or possibly an AGN,  because of its localised flux density (Fig.~\ref{fig3}) and steep spectral index.

\begin{figure}
  \includegraphics[width=80mm,angle=-90]{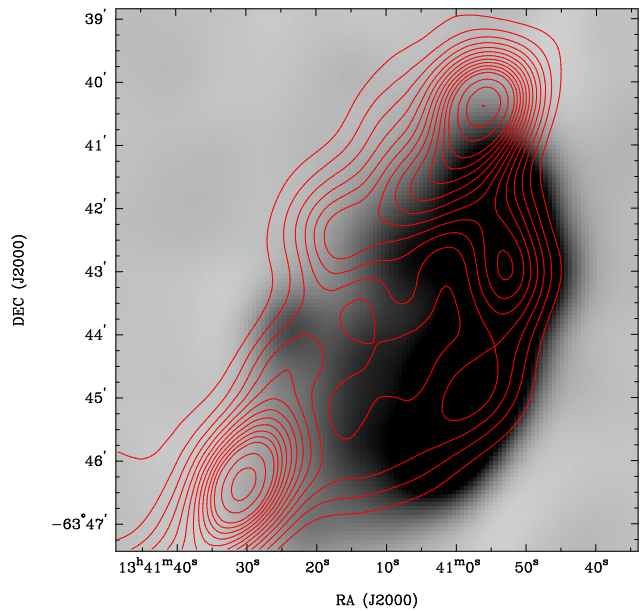}
  \caption{ Chandra X-ray image (maximum positional error \mbox{$\delta\alpha=3\farcs08$}, \mbox{$\delta\delta=1\farcs46$}) smoothed to match 20~cm beam overlaid with 20~cm contours. Contours at the following multiples of $3\sigma=2.1$~mJy: 1,2,3,\dots 12,14,16,18,20). }
   \label{fig5}
\end{figure}

The Chandra image (Fig.~\ref{fig5}) shows a strong correlation with the west side of the proposed SNR region. 
Such a one-sided, asymmetric morphology suggests a type II, Ib or Ic SN progenitor. If this were the case, then the east side of the `shell' seen in the image might be a radio jet emitting out of the opposite side of S3. However, despite there appearing to be no shock front X-ray emission on the eastern side of the SNR shell, the centre of the remnant still contains emission, appearing relatively brighter than the background noise. Additionally, no evidence can be found for an associated pulsar wind nebula or neutron star that would be expected from a type II, Ib or Ic SN progenitor. Prinz \& Becker (in prep.) consider the possibility of the X-ray emission originating from a galaxy cluster, but conclude that its X-ray spectral properties strongly favour an SNR nature. Given all this, and taking into account the likelihood of the different SNRs morphologies \citep{1999PhDT..........G}, we determine that this object is a shell SNR, slightly elongated in the north-south direction.

%FIGURE 5
%**********************************************************************************************************************************************************

%**********************************************************************************************************************************************************
\subsubsection{Angular Diameter}
\begin{figure}
\hbox{\hspace*{-2mm}\includegraphics[width=70mm,angle=-90]{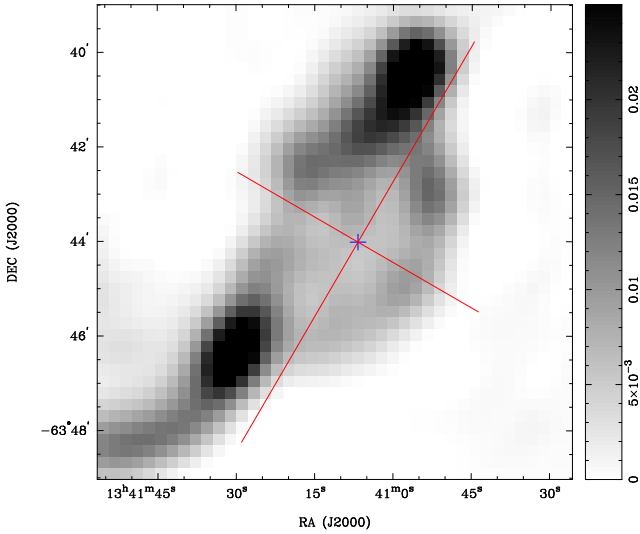}}
\caption{20-cm image of \SNR\ showing lines along which the line profiles in Fig.~\ref{fig7} are made.}
\label{fig6}
\end{figure}

\begin{figure}
\includegraphics[width=42mm]{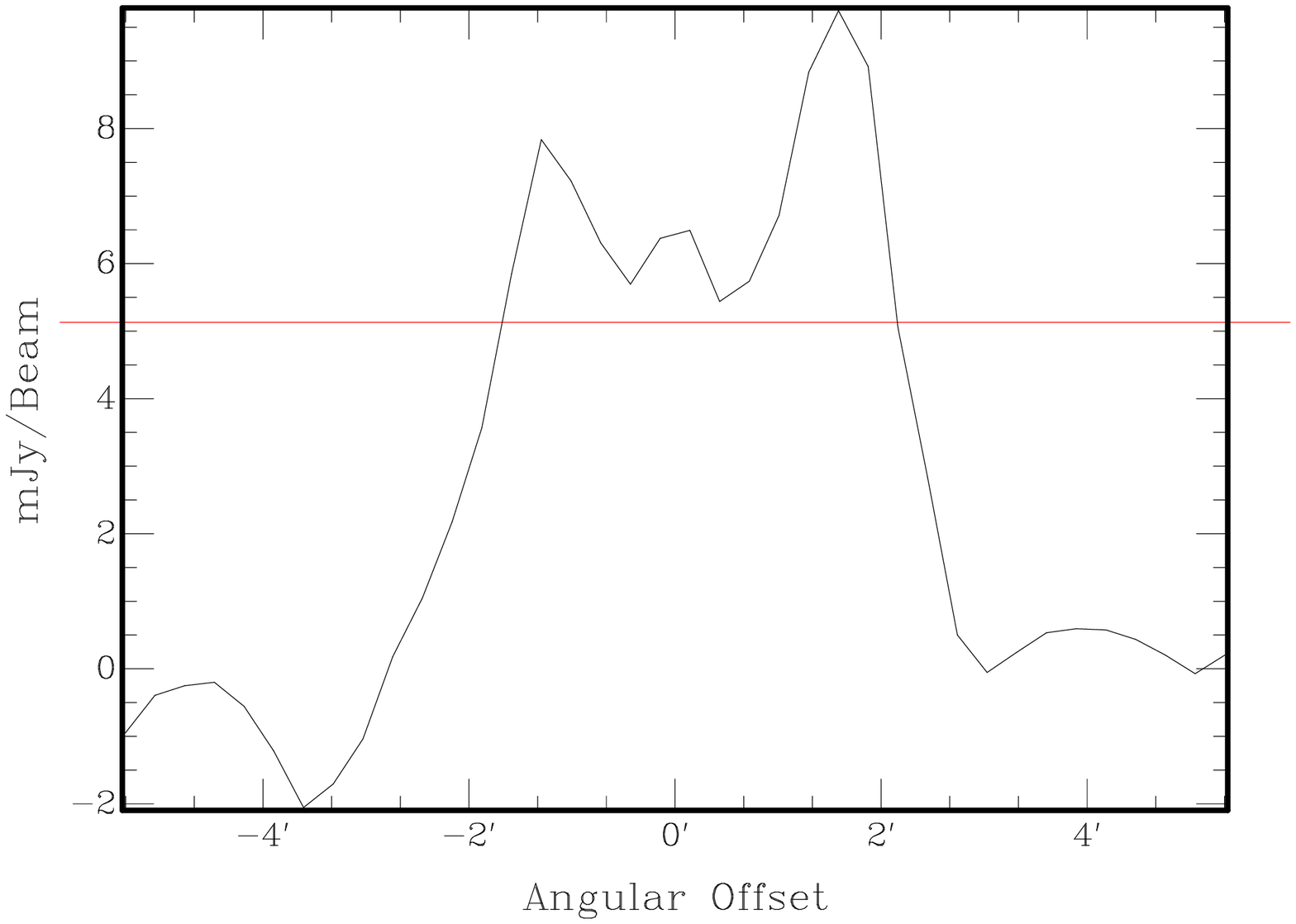}
\noindent\includegraphics[width=42mm]{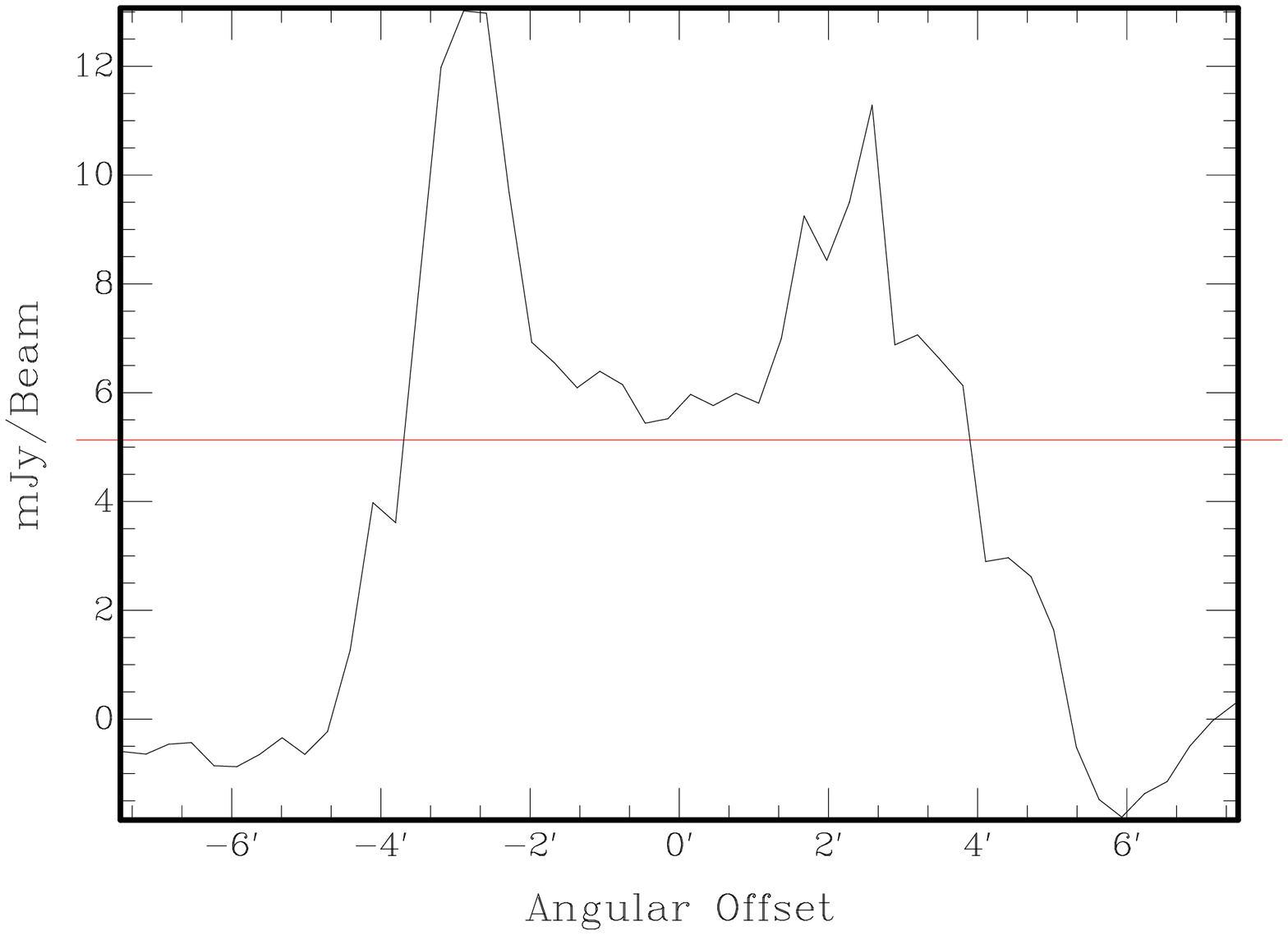}
\caption{Line profiles of the minor axis (left) and major axis (right). Red line indicates the $3\sigma$ noise level.}
\label{fig7}
\end{figure}

Profile plots of intensity against position, for the minor and major axes of the object (shown in Fig.~\ref{fig6}) are shown in Fig.~\ref{fig7}. From these, we determine the angular size of the SNR to be $4\arcmin\times8\arcmin\pm0\farcm5$ and the shell thickness (width of the peaks above $3\sigma$) to be $1\farcm00\pm0\farcm25$. 
This gives a volume filling factor $V_{Shell}/V_{SNR}$ of  $0.41\pm0.05$.

%FIGURE 6

%FIGURE 7
%**********************************************************************************************************************************************************

%**********************************************************************************************************************************************************
\subsection{Integrated Flux Density}
Table~\ref{Tab1} lists the estimated integrated flux density $S_\nu$ at the different frequencies. The integrated flux densities were estimated by including all the flux measurements corresponding to the remnant {\bf (i.e. within our SNR region as seen in Fig. 1.)}, which was determined by a contour at the $3\sigma$ level. The Parkes image suffers from confusion and the flux density measured includes the background sources. Additionally, it should be noted that the flux density at 13~cm was calculated based on an image constructed with a $uv$-range up to 4 k$\lambda$. We estimate that the error in flux densities is no more than $\pm10\%$.

%TABLE 1
\begin{table}
\begin{center}
\caption{Integrated flux densities of \SNR.}
\begin{tabular}{l c c c c c}
\hline
Radio & $\nu$ & $\lambda$ & $\sigma$ & Beam Size & $S_\nu$\\
Telescope & (MHz) & (cm) & (mJy~beam$^{-1}$) & (arcsec)  & (mJy)\\
\hline
MOST & \p0843 & 36 & \p01.5 & $45\times50$ & 257\\
ATCA & 1384 & 20 & \p00.7 & $65\times54$ & 215\\
ATCA & 2496 & 13 & \p00.5 & $41\times34$ & 124\\
arkes & 4850 & \p06 & 15.0 & $295\times295$ & \p084\\
\hline
\label{Tab1}
\end{tabular}
\end{center}
\end{table}
%**********************************************************************************************************************************************************

%**********************************************************************************************************************************************************
\subsection{Spectral Index}
\begin{figure}
\includegraphics[width=85mm,angle=-90]{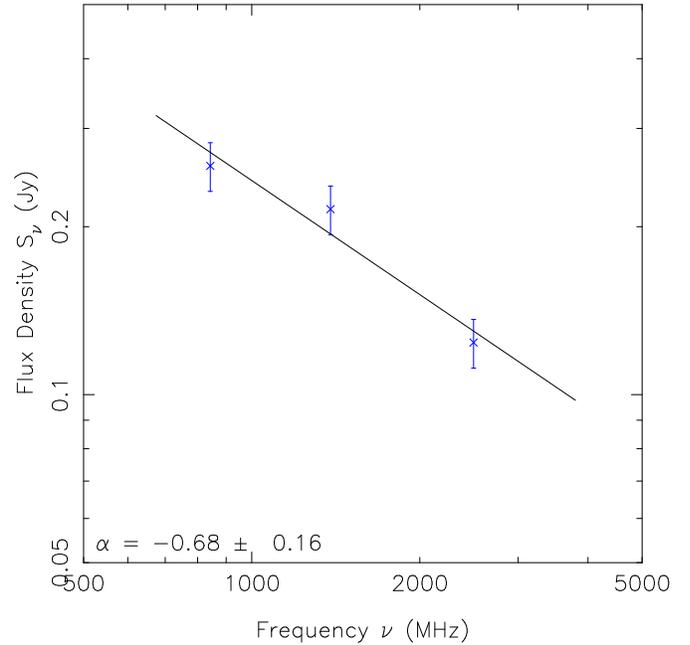}
\caption{Radio-continuum spectrum of \SNR\ between 36, 20 and 13~cm.}
\label{fig8}
\end{figure}

By fitting a power law to the measured integrated flux densities at wavelengths $\lambda$ = 36, 20 and 13~cm we created a three-point spectral index plot (Fig.~\ref{fig8}). The estimated spectral index is $\alpha = -0.68 \pm{0.16}$, indicating that the radio emission from \SNR\ is mostly non-thermal synchrotron radiation. This spectral index is consistent with a typical  SNR spectrum, with $-0.8$~\textless~$\alpha$~\textless~$-0.2$ \citep{1998A&AS..130..421F}.
Given this trend for the integrated flux density as a function~of frequency, we estimate the flux density at 1 GHz to be \mbox{$S_{\mathrm{1GHz}}\approx242$~mJy}.

%FIGURE 8

%FIGURE 9
\begin{figure}
  \hbox{\hspace*{-11mm}\includegraphics[width=75mm,angle=-90]{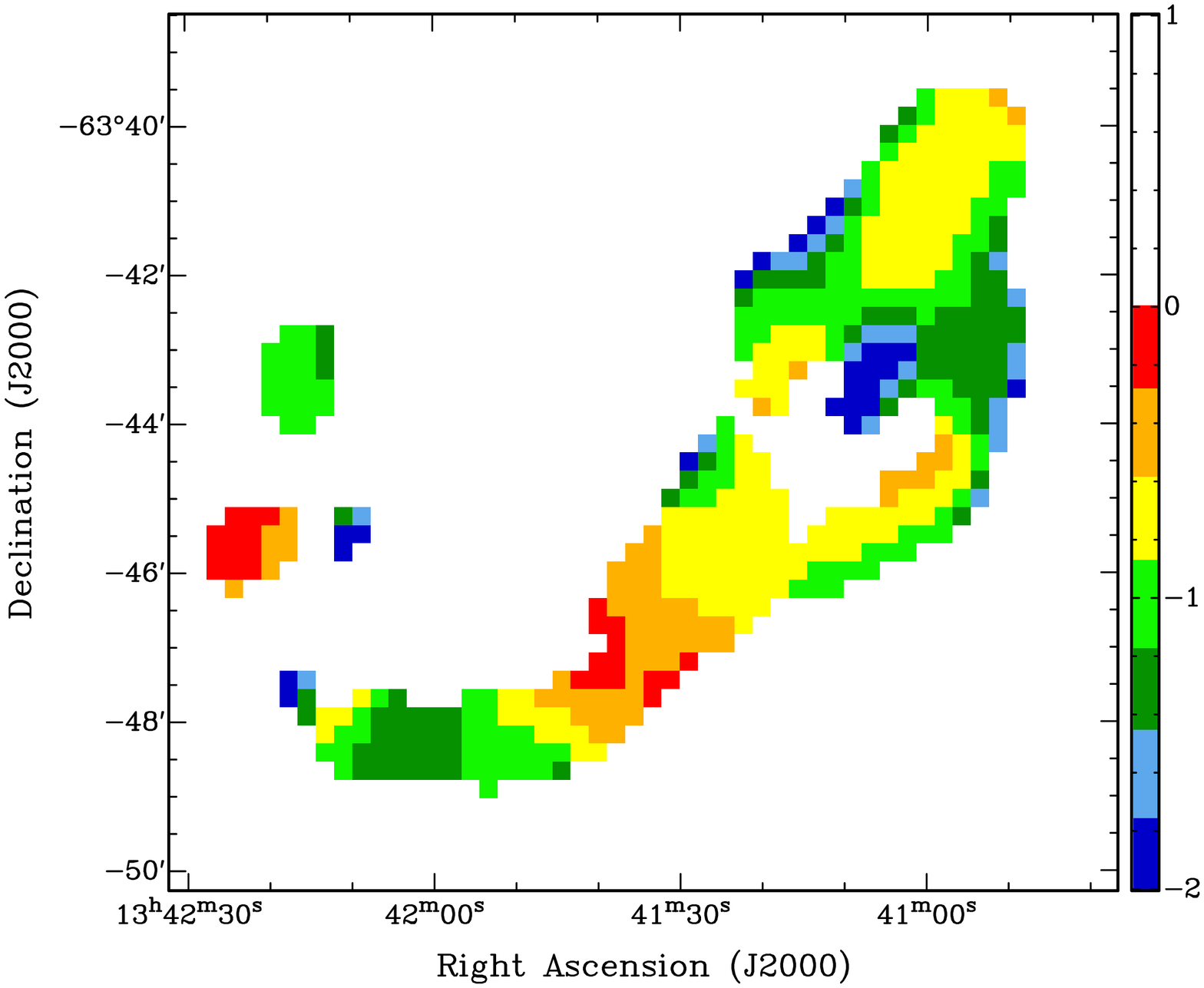}}
  \caption{3-point spectral distribution map: synthesised beam of \hbox{$65\farcs25\times53\farcs91$} at \hbox{P.A.$=-24\fdg1$}.}
   \label{fig9}
\end{figure}

In a similar fashion, we constructed the three-point spectral distribution map (36, 20 and 13~cm), shown in Fig.~\ref{fig9}. The map shows the region corresponding to the SNR having a spectral index ranging primarily between $-$0.5 \textless~$\alpha$~\textless~$-$1.5. The spectral index is quite steep at the edge of the shell of the SNR but becomes shallower toward the centre of the remnant. This distribution would be expected from a young to middle-aged SNR which accelerates particles efficiently at the strong shock front, producing strong synchrotron radiation.
%**********************************************************************************************************************************************************

%**********************************************************************************************************************************************************
\subsection{Distance Estimate}
Although controversial, the $\Sigma-D$ relation \citep{Danga,1998ApJ...504..761C,2004MNRAS.350..346A,2005A&A...435..437U,2010ApJ...719..950U} is the only method that can be used to estimate a distance from our data. It states that the surface brightness $\Sigma$ is related to the diameter $D$ by $\Sigma_\nu=AD^{-\beta}$, where $A$ is a constant dependent on the characteristics of the SN's initial explosion and the surrounding ISM, while $\beta$ is almost entirely independent of these. The theory surrounding the $\Sigma-D$ relation has evolved greatly since it was originally proposed, now incorporating many SNR cases, including that of bremsstrahlung radiation \citep{2005APh....23..577U}. \SNR\ exhibits features typical of a SNR in the adiabatic stage, such as bright X-ray emission, a negative spectral index and significant linear polarisation. Since precise details about the local environment of \SNR\ are unknown, we estimated a distance to the SNR using the new Galactic $\Sigma-D$ relation \citep{PavApJinpress}  which is calibrated with 50 Galactic shell-like SNRs with independently determined distances using the orthogonal fitting procedure.

We estimate the surface brightness at 1~GHz to be \mbox{$\Sigma_{\mathrm{1GHz}}$= 1.1$\times$10$^{-21}$ W m$^{-2}$ Hz$^{-1}$ sr$^{-1}$} which gives an average diameter of $D=34\pm$19~\mbox{pc} and thus an average distance of $d=19\pm11~\mbox{kpc}$, which would indicate that \SNR\ is on the far side of the Galaxy. This average distance comes from the average of the major and minor axes, while the uncertainties arise from the upper and lower limits for $A$. At this distance, \SNR\ would lie $\sim 500$pc above the Galactic plane, with linear dimensions of $23\times45\pm3\rmn{ pc}$ and shell thickness $\sim 5.5$ pc.

Placing \SNR\ on the \citet{2004A&A...427..525B} surface brightness-diameter diagram (at 1 GHz) suggests that \SNR\ is in the early adiabatic phase of evolution, expanding into an extremely low-density environment with an initial energy somewhat lower than the canonical SN energy of $10^{51}$ergs. The suggested low-density environment is consistent with our estimated position of the SNR, 0.5 kpc out of the Galactic plane. 
%**********************************************************************************************************************************************************

%**********************************************************************************************************************************************************
\subsection{Polarimetry}

%FIGURE 10
\begin{figure}
\begin{center}
\includegraphics[width=60mm,angle=-90]{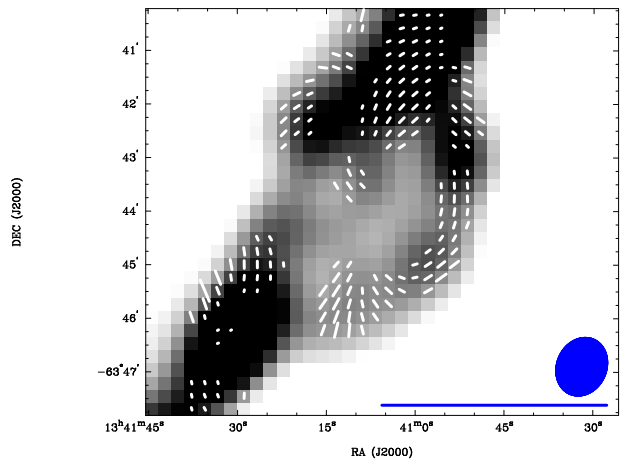}
\end{center}
\caption{20 cm ATCA observations of \SNR. The ellipse in the lower right corner of the image represents the synthesised beam width of \hbox{$65\farcs25\times53\farcs91$} at \hbox{P.A.$=-24\fdg1$}. The length of the vectors represent the fractional polarised intensity at each pixel position, and their orientations indicate the mean position angle of the electric field (averaged over the observing bandwidth, not corrected for any Faraday rotation). The blue line below the beam ellipse represents the length of a fractional polarisation vector of 100\%. The maximum fractional polarisation is \mbox{$10\%\pm1\%$}.}
\label{fig10}
\end{figure}

Since the ATCA observations recorded Stokes parameters $Q$, $U$ and $V$, in addition to total intensity $I$, we were able to determine the polarisation of the SNR at 20~cm. Fig.~\ref{fig10} shows the regions polarised above a level of 3$\sigma$. The electric field vectors are parallel with the shell of the SNR around most of the circumference of the SNR, particularly along its western side. This arrangement is expected as the SNR's particles are experiencing the greatest acceleration and most efficiently producing synchrotron radiation. The maximum fractional polarisation is is estimated to be P=10$\pm1\%$. The polarisation vectors overlaying background source S4 are probably unreliable, due to possible confusion between the SNR and background source. 
No reliable polarisation was detected at 13~cm. This might indicate significant depolarisation in the remnant, but the polarimetric response of the ATCA is poor at 13~cm.

We used the new equipartion formula derived by \citet{0004-637X-746-1-79} based on the diffusive shock acceleration (DSA) theory of \citet{1978MNRAS.182..443B} to estimate a magnetic field strength: This formula is particularly relevant to magnetic field estimation in SNRs, and yields magnetic field strengths between those given by the classical equipartition \citep{1970ranp.book.....P} and revised equipartition \citep{2005AN....326..414B} methods. We estimate the magnetic field strength, $B \approx 29~\mu\mbox{G}$, and the minimum total energy of the synchrotron radiation, $E_{min} \approx 2\times10^{49}~\mbox{ergs}$ (see \citet{0004-637X-746-1-79}; and corresponding online ``calculator" \footnote{http://poincare.matf.bg.ac.rs/$\sim$arbo/eqp/}).

The value of 29 $\mu$G indicates that this SNR's magnetic field is not only influenced by the compression of the ISM magnetic field ($\sim$1 $\mu$G in a low density environment), but is also (moderately) amplified by DSA effects (see \citet{2004MNRAS.353..550B}). The relatively young estimated age of this SNR, somewhat steep spectral index (\citet{2011MNRAS.418.1208B}), and moderately amplified magnetic field, constitute the standard description of an SNR in the early Sedov phase of evolution.
%**********************************************************************************************************************************************************

%**********************************************************************************************************************************************************
\section{Conclusions}
Analysis of our new radio-continuum images of the galactic radio source \SNR, produced from from $\lambda=$ 20 and 13~cm ATCA archival data
confirms that the object is indeed a {\em bona fide} SNR. The overall spectral index of \SNR\ is \mbox{$\alpha=-0.68\pm{0.16}$}, with the emission \mbox{10$\pm1\%$ linearly polarised}.
These properties indicate that it is likely to be young to middle-aged, with the majority of its radio emission being non-thermal, due to the strong shock wave driving the shell and accelerating particles creating synchrotron radiation. We suggest \SNR\ is in the early adiabatic expansion stage, due to its significant X-ray brightness, steep non-thermal spectral index and linear polarisation. We estimate the magnetic field strength, \mbox{$B \approx 29~\mu\mbox{G}$},  which is consistent with the SNR being at this evolutionary stage.

Taking these new ATCA radio-continuum images in conjunction with the high resolution Chandra X-ray image, we find that \SNR\ is likely to be of an elongated shell-like morphology, with the X-ray emission appearing stronger on the western side of the shell. No evidence for an associated compact central object was found.

We have used the new Galactic $\Sigma-D$ relation  of \citet{PavApJinpress} to estimate the dimensions of the source and thus its distance: \SNR\ was found to have an average diameter of \mbox{34$\pm$19 pc} and a distance of \mbox{19$\pm$11 kpc}.

Our radio images could be greatly improved by further ATCA observations with better parallactic angle coverage and higher resolution. Additional X-ray spectral observations with the XMM-Newton satellite would also yield new insights into the elemental abundances and energies of \SNR.
%**********************************************************************************************************************************************************

%**********************************************************************************************************************************************************
\section{Acknowledgments}
We used the \textsc{karma} software package developed by the ATNF. The Australia Telescope Compact Array and Parkes radio telescope are part of the Australia Telescope National Facility which is funded by the Commonwealth of Australia for operation as a National Facility managed by CSIRO. This paper includes archived data obtained through the Australia Telescope Online Archive (http://atoa.atnf.csiro.au). This research is supported by the Ministry of Education and Science of the Republic of Serbia through project No. 176005. 
%**********************************************************************************************************************************************************

%**********************************************************************************************************************************************************
\bibliographystyle{mn2e}
\bibliography{G308-Paper01}
\label{lastpage}
\end{document}